\documentclass[aps,prb,showkeys,twocolumn,showpacs,11pt]{revtex4-1}
\usepackage{graphics,graphicx}
\usepackage{color}
\usepackage{amsbsy,amssymb}
\usepackage{amsmath}
\usepackage{epsfig}
\usepackage{color}
\usepackage{textcomp}
\begin{document}
\title{Graphene composites: The materials for the future}
%%%%%%%%%%%%%%%%%%%%%%%%%%%%%%%%%%%%%%%%%%%%%%%%%%%%%%%%%%%%%%%%%%%%%%%%%%%%%%%%%%%%%%%%%%%%

%%%%%%%%%%%%%%%%%%%%%%%%%%%%%%%%%%%%%%%%%%%%%%%%%%%%%%%%%%%%%%%%%%%%%%%%%%%%%%%%%%%%%%%%%%%
\author{Sreemanta Mitra$^{1,2}$}
\email[]{sreemanta85@gmail.com}
%\author{Oindrila Mondal$^{3}$}
%\author{Dhriti ranjan Saha$^{1}$}
\author{Sourish Banerjee$^{2}$}
\author{Anindya Datta $^{3}$}
%\email[]{anindya$_$datta@yahoo.com}

\author{Dipankar Chakravorty$^{1,\dag}$}
\email[]{mlsdc@iacs.res.in}
%%%%%%%%%%%%%%%%%%%%%%%%%%%%%%%%%%%%%%%%%%%%%%%%%%%%%%%%%%
%\author{Sreemanta Mitra$^{1,2}$,Amrita Mandal$^{1,2}$,Anindya Datta$^{1,3}$,Sourish Banerjee$^{2}$ and Dipankar Chakravorty$^{1,\dag}$}
\affiliation{
$^{1}$
 MLS Prof.of Physics' Unit,Indian Association for the Cultivation of Science, Kolkata-700032, India.\\ }
\affiliation{
$^{2}$
Department of Physics, University of Calcutta, Kolkata-700009, India.\\}
%\affiliation{
%$^{3}$ Department of Physics, M.U.C. Woman's College, Burdwan, India.\\}

\affiliation{
$^{3}$
University School of Basic and Applied Science (USBAS),Guru Govind Singh Indraprastha University,New Delhi, India\\}
%%%%%%%%%%%%%%%%%%%%%%%%%%%%%%%%%%%%%%%%%%%%%%%%%%%%%%%%%%%%%%%%%%%%%%%%%%%%%%%%%%%%%%%%%%%%%%
%\begin{document}
%%%%%%%%%%%%%%%%%%%%%%%%%%%%%%%%%%% Abstract %%%%%%%%%%%%%%%%%%%%%%%%%%%%%%%%%%%%%%%%%%%%%%%%%%%%%%%
\begin{abstract}
 The exotic physical properties of graphene have led to intense research activities on the 
 synthesis and characterization of graphene composites during the last decade. In this article the
 methods developed for preparation of such materials and the different application areas are reviewed.
 The composites discussed are of two types, viz; graphene/polymer and inorganic/ graphene.
 The techniques of ex-situ hybridization and in-situ hybridization have been pointed out. Some of the 
 application areas are batteries and ultracapacitor for energy storage and fuel cell and solar cell for 
 energy generation and some of the possible future directions of research have been discussed. 
 
\end{abstract}

\maketitle
%%%%%%%%%%%%%%%%%%%%%%%%%%%%%%%%%%%%%%%%%%%%%%%%%%%%%%%%%%%%%%%%%%%%%%%%%%%%%%%%%%%%%%%%%%%%%%%%%%
\section{Introduction}\label{sec.1}

Since its discovery \cite{geimsc} the first two dimensional material, graphene,
which consists of $sp^{2}$ bonded carbon atoms in honeycomb network,has been in
the limelight of research in condensed matter physics. Synthesis of graphene belied
the argument of Peirls and Landau, that two dimensional materials could not sustain 
thermodynamic fluctuation, and hence would be unstable.\cite{novopnas}
The unique properties of graphene arise due to the long range $ \pi $ conjugation 
which results in high value of Young's modulus ($\sim 1 TPa$)\cite{science321}
high thermal conductivity ($\sim 5000 Wm^{-1}K^{-1}$) \cite{nl8} and high mobility 
of charge carriers($\sim 200000 cm^{2}V^{-1}s^{-1}$)\cite{ssc146}. These intriguing 
properties of graphene, tempted researchers to fabricate composites with the same.
A nanocomposite is a combination of multiple materials, out of which one has a nanometer 
dimension, and the property may reflect that of a weighted average of the components or 
a completely new one. Previously, another well known carbon based material viz; carbon 
nanotube (CNT) was used to synthesize nanocomposites, which showed interesting properties. 
Graphene has  quite a few properties similar to those of CNTs. It also shows
some distinctly different behavior, like quantum Hall effect both anomalous and fractional,
due to the ballistic nature of its charge carriers.\cite{geimsc, novonat, geimnatphy}
A number of graphene based nanocomposites with inorganic matrices
\cite{wujpcc,yinsml,zhoujpcc,muzynskijpcc}, organic crystals \cite{wangacsn1,hanadv},
bio-materials\cite{wangjacs1},polymers (both conducting and non conducting)
\cite{qiac,qisml,yanglm,ramnn}have been studied for various applications, such as supercapacitors,
photocatalysis,sensors etc. In this article, we review the work carried out on graphene
based composites.
%%%%%%%%%%%%%%%%%%%%%%%%%%%%%%%%%%%%%%%%%%%%%%%%%%%%%%%%%%%%%%%%%%%%%%%%%%%%%%%%%%%%%%%%%

\section{Properties}\label{sec.2}
Before starting the discussion on graphene based composites, we briefly describe 
the electronic properties of graphene, as these are the  important  
aspects which determine its applications. In 1946, P.R.Wallace was the first to show theoretically,
that graphene is semi-metallic in nature, and the low energy excitations are similar to those of 
Dirac fermions. The chemical potential crosses exactly the Dirac point and hence gives rise to a
linear dispersion spectrum\cite{wallacepr}. The hexagonally arranged carbon atoms in graphene can be visualized 
as a triangular lattice with two atoms per unit cell, as a basis. The s, $p_{x}$, and $p_{y}$
orbitals of a carbon atom make $\sigma$ bond with the neighboring carbon atom. The 
hybridization of $p_{z}$ orbitals of neighboring carbon atoms results in $\pi$ and $\pi^{*}$ bands
which cross at K and K$^{'}$ point and give a linear dispersion. By first nearest neighbor
tight binding model,the E-K relation of these $\pi$ electrons has been shown to be given by,
\cite{castronetormp}
%%%%%%%%%%%%%%%%%%%%%%%%%%%%%%%%%%%%%%
\begin{equation}
E^{\pm}=\pm \gamma \sqrt{1+4\cos{\frac{\sqrt{3}k_{x}a}{2}}\cos{\frac{k_{y}a}{2}}+4{\cos}^{2}{\frac{k_{y}a}{2}}}
\end{equation}
%%%%%%%%%%%%%%%%%%%%%%%%%%%%%%%%%%%%%%%
where, $\gamma$ represents a positive constant, known as transfer integral, and is the matrix element
between the $\pi$ orbitals of the neighboring carbon atoms. The positive and negative signs in 
equation (1) signify fully empty anti-bonding $\pi^{*}$ and fully filled bonding $\pi$ band 
respectively.Up to first approximation, for small k, the equation (1) is reduced to,
%%%%%%%%%%%%%%%%%%%%%%%%%%%%%%%%%%%%%%%
\begin{equation}
 E^{\pm}(\vec{k})=\pm \gamma_{0}\vec{k}
\end{equation}
%%%%%%%%%%%%%%%%%%%%%%%%%%%%5
$\vec{k}$ is measured with respect to the K points and 
$\gamma_{0}=\hslash v_{f}=\frac{\sqrt{3}a\gamma}{2}$, and $v_{f}$ is the Fermi group velocity.
This linear dispersion relation in graphene, allows one to leave $Schr\ddot{o}dinger$'s equation
as used in most condensed matter systems, and consider Dirac equation instead to demonstrate
the particle behavior. The idea has a direct consequence to replace the effective mass by 
cyclotron mass and it has also been found that $m^{*}$ is directly proportional to the 
electron density\cite{novonat}.

%%%%%%%%%%%%%%%%%%%%%%%%%%%%%%%%%%%%%%%%%%%%%%%%%%%%%%%%%%%%%%%%%%%%%%%%%%%%%%%%%%%%%%%%%%%%%%%
\section{Graphene in composite}\label{sec.3}
The method of preparation of graphene as developed by Geim and Novoselov\cite{geimsc}, viz;
\textquoteleft The scotch tape method \textquoteright 
produces high quality (as observed from Raman spectroscopy and AFM images ) graphene
but this is not suitable for large scale production. Hence, the technique could not be used for 
composite synthesis. Apart from this Nobel winning method, five other widely used methods have been
reported for the synthesis of graphene, namely, (a) chemical vapor deposition\cite{reinanl,jmc21}, 
(b)epitaxial growth\cite{herrssc,sutternm,shivnl}, (c)chemical reduction of graphite derivative
\cite{stancarb,parknn}, (d) ``Unzipping'' of CNTs\cite{kosynkinnat,jiaonat}, (e) organic synthesis
technique\cite{yanjacs}. Out of these methods, the chemical reduction of graphite 
derivatives has been employed mostly for synthesizing composites based on graphene.This strategy has
become popular, since it not only yields large amount of chemically modified graphene, but also can be 
used to functionalize it suitably for practical application.
\par 
At present most of the composites based on graphene use reduced graphene oxide (RGO),which is obtained
by the exfoliation and chemical reduction of graphite oxide (GO), a graphite derivative.
Of several methods prescribed,modified Hummers' method has mostly been used to synthesize
graphite oxide by treating graphite with strong mineral acids. Though the structure of graphite oxide
is a point of debate till date, oxidation of graphite will destroy the $sp^{2}$ hybridization
and increase the interlayer separation. As a result, the Van-der Waal's force between the layers gets
much weaker, and it becomes easy to exfoliate. The chemical reduction of GO to RGO can be done by 
using several reducing agents; viz, hydrazine hydrate \cite{stancarb} sodium borohydride 
\cite{shencm,linn,advfm19},hydroquinone\cite{wangjpcc}etc. 
Hydrazine has mostly been used as the reducing agent.
Many environment friendly chemicals have been used to reduce
GO to form RGO, such as alcohols, sodium citrate, tea, solar radiation etc.
\cite{dreyerjmc,zhangnr,wangacsami,varrlajmc}.
   \subsection{Classifications}
The synthesis of first ever graphene based composites, 
viz; graphene/polystyrene was reported in 2006, by 
R.S.Ruoff and co-workers\cite{stanknat}.
Their main observation was low percolation threshold, which they had computed to be 
0.1 vol\% of graphene for the composite. The observation have been given in Fig. 1.
%%%%%%%%%%%%%%%%%%%%%%%%%%%%%%%%%%%%%%%%%%%%%%%%%%%%%%%%%%%%%%%%%%%%%%%%%%%%%
\begin{figure}
\begin{center}
 \includegraphics[width=8.5cm]{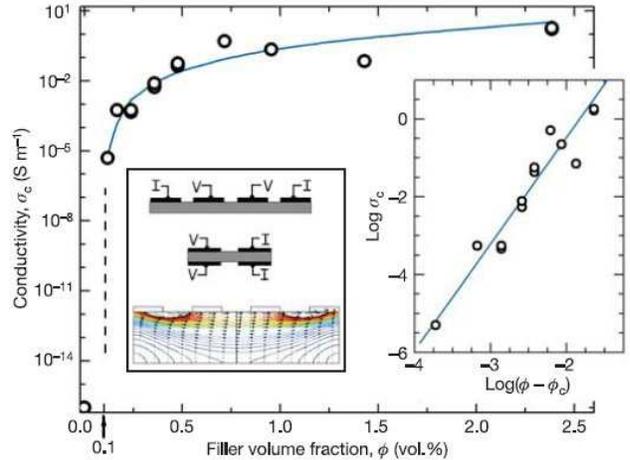}
\end{center}
\caption{Electrical conductivity of the graphene/polystyrene composites
as a function of filler volume fraction. Reprinted with permission from \cite{stanknat}, Nature Publishing Group.} 
\end{figure}
%%%%%%%%%%%%%%%%%%%%%%%%%%%%%%%%%%%%%%%%%%%%%%%%%%%%%%%%%%%%%%%%%%%%%%%%%%%%%%%%%%%

This work attracted considerable attention which led to
the further development of different composites with graphene. Broadly, we can classify graphene 
composites under two major groups viz;(a)graphene/polymer composite (b)graphene/inorganic material
composites.
\par
We  briefly discuss below the work carried out on the above two categories of graphene composites.
%%%%%%%%%%%%%%%%%%%%%%%%%%%%%%%%%%%%%%%%%%%%%%%%%%%%%%%%%%%%%%%
    \subsubsection{Graphene/Polymer composites}
      Graphene has been successfully used as a filler material, similar to the CNT in polymers. These 
can be sub categorized in to two parts,(i)graphene/non-conducting polymer composite and 
(ii)graphene/conducting polymer composites.\\
%%%%%%%%%%%%%%%%%%%%%%%%%%%%%%%%%%%%%%%%%%%%%%%%
\begin{figure}
 \begin{center}
  \includegraphics[width=8.5cm]{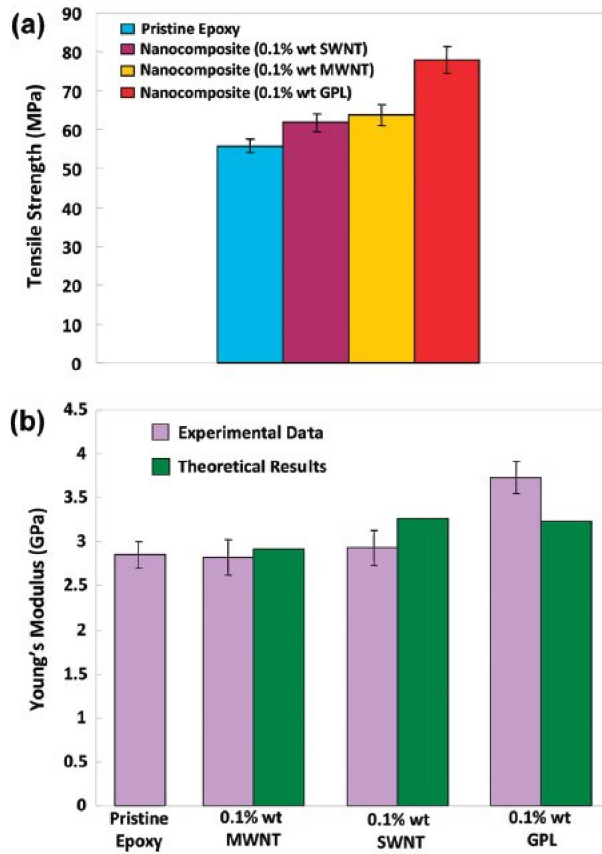}
 \end{center}
\caption{Uniaxial tensile testing. (a) Ultimate tensile strength for
the baseline epoxy and GPL/epoxy, MWNT/epoxy, and SWNT/epoxy
nanocomposites. The weight fraction of nanofillers for all of the
nanocomposite samples tested was fixed at 0.1\%. (b) Young’s
modulus of nanocomposite samples with 0.1\% weight of GPL,
0.1\% weight of SWNT, and 0.1\% weight of MWNT is compared
with the pristine (i.e., unfilled) epoxy matrix. Reprinted with permission from \cite{rafieeacsn}, American Chemical Society.}
\end{figure}
%%%%%%%%%%%%%%%%%%%%%%%%%%%%%%%%%%%%%%%%%%%%%%%%%%%%%%%%%%%%%%%%%%

In the polymer matrix; graphene fillers of different volume fractions can  substantially improve
the properties, e.g. electrical,thermal {\&} mechanical. Due to their high aspect ratio 
and large interface graphene/polymer composites show much enhanced properties at low concentration
than CNT/polymer composites\cite{rafieeacsn,liangadvfm}. The summarized results have been described in Fig. 2. 
       \par
The most simple and straightforward method to fabricate polymer composites is solution mixing.
To achieve good dispersion, the compatibility of the polymer and filler (graphene in this case)
in the required solvent is important. GO is easily dispersed in water, due to its oxygen
functionality, but RGO has very low solubility, hence RGO in water was regularly sonicated in
water in an ultrasonic bath, to achieve good dispersion before mixing it with different aqueous
solutions of polymers, e.g. (poly)methyl methacrylate (PMMA)\cite{ramjpsb}, (poly)aniline (PANI)
\cite{wuacsn},(poly)urethene\cite{caint} and (poly)vinyl alcohol(PVA)\cite{mitrajpcc}.\\
An unavoidable situation of re-stacking and aggregation of graphene sheets develops in the above method of
synthesizing graphene/polymer composites. This could be taken care of  by surface functionalization 
of the RGO before mixing with polymers. \\
Another approach of synthesis is in-situ polymerization technique\cite{yujpcc,xiaojmc,rafieesml,
zhoucc,zhangcm}.\\
Alternatively, by the method of in-situ anodic polymerization\cite{wangacsn2} graphene/PANI composite
was fabricated. Apart from this, graphene was also included as fillers in different polymer matrices
such as silicone\cite{verdejojmc} via this method.
\par
Enhancement of some properties or obtaining a completely new property is the primary objective of 
fabricating composites. Graphene/polymer composites attain these objectives satisfactorily.\\
The electrical conductivity of different polymers filled with graphene have shown enormous improvement.
Percolation threshold has been found to be very low in different types of insulating polymers as matrices
in composites, e.g. 0.15 volume{\%} RGO in (poly)vinyl chloride/vinyl acetate co-polymer
composite \cite{weicarb}, 0.47{\%} filler in graphene/PET composites\cite{zhangplym}. This is
due to graphene's large surface area and $\pi$ conjugated 2D surface. Also filler/matrix interaction
mediated by surface functional groups of graphene has a moderate role to play in this operation.
Along with the electrical conductivity, the thermal conductivity of the polymer matrices with 
graphene as fillers have been shown to increase substantially. The high room temperature (RT) thermal
conductivity of single layer graphene,viz;($\sim 5000 Wm^{-1}K^{-1}$) showed a prominent effect when
only 0.25 volume{\%} RGO in silicon foam matrix, increased its thermal conductivity by 6{\%}
\cite{verdejojmc}. This increase of thermal conductivity,surpasses different conventional 
fillers like Ag, which need around 50{\%} loading to achieve the same value.
Apart from these, mechanical properties and thermal stability of different polymers with graphene 
fillers have shown increase compared to that of the polymers themselves. Being the strongest material
till date, graphene's incorporation in isocyanate PU matrix increased the Young's modulus and hardness
by 900{\%} and 327{\%} respectively\cite{caint}.
Rao et.al. have shown \cite{cnrpnas1} that,SWNT/PVA composites exhibit superior mechanical proeperties
as compared to that of graphene/PVA composites. They have reported that composites comprising of two 
different nanocarbon  (graphene/SWNT), (SWNT/Nanodiamond) and (Nanodiamond/graphene) reinforced in polymer 
show a large enhancement of the mechanical property. Functional polymers' another important property, viz;
thermal stability was also improved by incorporating graphene (as it has a superior properties)
in the polymer. To name a few,the glass transition temperature (temperature above which the chains of a
thermoplastic polymer begin to flow) of 0.05 wt.{\%} loaded graphene in PMMA increased by 30K and 
1 wt.{\%} loaded graphene in (poly)acrylonitrile by 40K\cite{ramnn}. Graphene/elastomer composite
showed better thermal stability viz; an increase of the degradation temperature by 55K 
(0.25{\%}RGO/silicone)\cite{verdejojmc}and 10K (0.5{\%} exfoliated graphite/PLA)\cite{kimjpsb}.
The strong interlayer cohesive energy and surface area along with strong filler/matrix interaction 
have been shown to be responsible for this enhancement.\\
The highly hydrophilic PVA became highly hydrophobic by incorporation of graphene fillers in it
\cite{wangplyi}
\par
In addition to the polymer composites with randomly distributed graphene as a filler in polymer
matrices, layered graphene/polymer composite was also fabricated for specific application purposes
e.g. photovoltaic thin films\cite{xucarb}.Employing Langmuir-Blodgett (LB) technique, GO sheets 
were deposited on (poly)allylamine hydrochloride multilayer, which eventually, showed enhancement
of directional elastic modulus, by an order of magnitude\cite{kulacsn}. For photovoltaic applications
(poly)3-hexylthiophene(P3H7)/phenyl-C61-butyric acid (PCBM)/GO were deposited on the ITO coated 
substrate in which GO acted as a hole transport segment\cite{liacsn}.

%%%%%%%%%%%%%%%%%%%%%%%%%%%%%%%%%%%%%%%%%%%%%%%%%%%%%%%%%%%%%%%%%%%%%%%

\subsubsection{Graphene/Inorganic composite}
    Nanoparticles or nanostructures of different shapes and sizes have got increasing attention from
research point of view during the last decade, because of their interesting properties.
Several nanoparticles of metals, metal oxides,and other inorganic compounds have been made into 
a composite structure with graphene to further enhance their properties. The materials included
in the list are Au, Ag \cite{zhoujpcc}, Ni\cite{wangjacs2}, Cu\cite{hassanjmc}, Ru, Rh\cite{marcarb}
$TiO_{2}$ \cite{liuadvfm}, ZnO\cite{yinsml}, $MnO_{2}$\cite{chenacsn}, $Co_{3}O_{4}$\cite{wuacsn},
NiO\cite{sonacsn}, $Fe_{3}O_{4}$\cite{shenjpcc} etc. The scheme for preparing $MnO_{2}$/graphene composite
has been shown schematically in fig. 3. 
%%%%%%%%%%%%%%%%%%%%%%%%%%%%%%%%%%%%%%%%%%%%%%%%%%%%%%%%%%%%%%%%%%%%%%%%%%%%%%%%%%%%%%%%%%%%%%%%%
\begin{figure}
 \begin{center}
  \includegraphics[width=8.5cm]{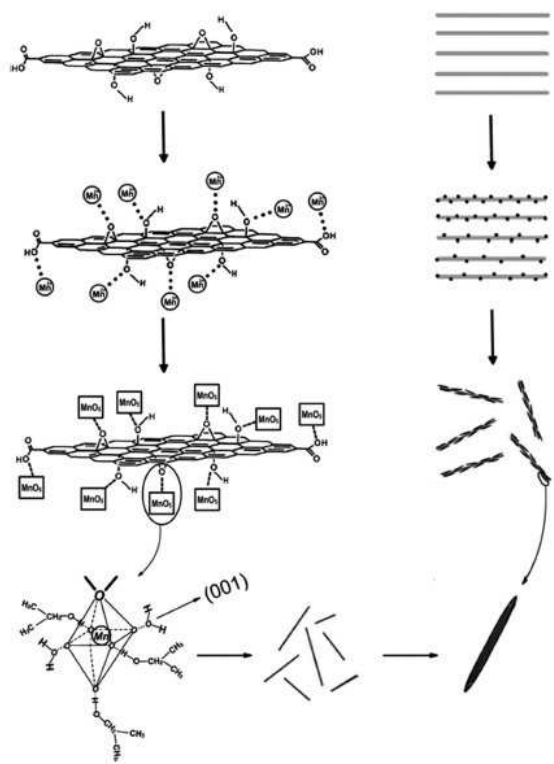}
 \end{center}
\caption{Scheme for preparing graphene/$MnO_{2}$ composite. Reprinted with permission from \cite{chenacsn}, American Chemical Society.}
\end{figure}
%%%%%%%%%%%%%%%%%%%%%%%%%%%%%%%%%%%%%%%%%%%%%%%%%%%%%%%%%%%%%%%%%%%%%%%%%%%%%%%%%%%%%%%%5
\par
In general, the synthetic methods for graphene/inorganic nanostructures can be classified into  
two categories, (a) ex-situ hybridization and (b)in-situ hybridization.
\par
\textbf{(a) Ex-situ hybridization: assembly on graphene}\\
Ex-situ hybridization technique is probably much more simplistic than the in-situ one. In general,
graphene and a pre synthesized or commercially available nanocrystals are mixed in solutions.
For better mixing, surface modification of either or both of the nanocrystals and the graphene was done. 
However, this functionalization is not a regular procedure. In this process, the $\pi$-$\pi$
stacking via some linking molecule is one of the mechanisms involved. The $\pi$-$\pi$ stacking
serves here as an adhesive layer to absorb the nanoparticle. Out of the several examples,
2-mercaptopyridine modified gold nanoparticles\cite{huangns} or benzyl mercaptan capped CdS 
nanoparticles\cite{fengnt} were successfully added to graphene (RGO) which have several applications
e.g. in catalysis, surface enhanced Raman spectroscopy (SERS) etc. Alternatively, modified or 
functionalized graphene has also been used for attaching nanoparticles to it. For example,
adhesive polymer nafion coated RGO has been used to prepare $RGO/TiO_{2}$ composites\cite{sunapl},
or bovine serum albumin (a bio-polymer) has been employed to modify RGO for the fabrication of 
silver, palladium, or platinum nanoparticle/graphene composites\cite{liujacs}.Apart from $\pi$-$\pi$
stacking (a non-covalent interaction) covalent interaction has also been used to fabricate composites.
For this GO, instead of RGO was used, since it (GO) has a large number of oxygen containing groups
which in turn can be used to link with other functional groups. In this approach, 
tetraethyl orthosilicate and (3-aminopropyl) triethoxysilane treated $Fe_{3}O_{4}$ was reacted 
with GO, in presence of 1-ethyl-3-(3-dimethylaminopropyl) carbodiimide and N-hydroxysuccinnimide
to form $Fe_{3}O_{4}$/GO composite. The amino groups introduced on functionalized $Fe_{3}O_{4}$
reacted with carboxylic groups of GO\cite{hecarb}. Amide bonds of 4-aminothiophenol functionalized
CdS nanoparticles were used to link with acylated GO sheets to form CdS/GO composite\cite{phamnt}.
By this covalent interaction, aminated CNTs and acid chloride activated GO were linked to form 
CNT/GO composite\cite{kimcarb}. Also by the method of Fisher esterification between hydroxyl groups
on fullerenes, GO/fullerene composite was synthesized\cite{zhangjmc}.
Electrostatic interactions were also used to prepare composites of graphene and inorganic 
nanoparticles. Due to the ionization of the oxygen containing functional groups on GO or RGO, they 
were negatively charged, and this was exploited to assemble positively charged inorganic 
nanoparticles on them via electrostatic interaction. $RGO/Fe_{3}O_{4}$ composite was synthesized by 
this route\cite{zhucpc}. In the case of negatively charged nanoparticles, the RGO/GO surface had 
to be changed to a positively charged configuration for the electrostatic interaction to take place.
For this the RGO was coated with cationic polyelectrolytes e.g. PQ11 treated RGO was attached with 
negatively charged silver (Ag) nanoparticles\cite{liumm}, and (poly) diallyldimethyl ammonium chloride
(PDDA) coated RGO acted as host for citrate capped gold nanoparticles, and thioglycolic acid 
functionalized CdSe quantum dots\cite{fanglm,liadvfm}.\\
Another exciting method of synthesizing graphene/inorganic nanoparticle composite was layer by layer 
self assembly, where one graphene layer was sandwiched between two layers of nanoparticle assembly.
In comparison to other known methods of multilayer fabrication, where composites were prepared
in advance, this technique allowed one to synthesize two components simultaneously, and the thickness
could be controlled with high precision in the nanoscale\cite{xiangacsami}. The electrostatic
interaction between the alternating layers of graphene and other materials was used to fabricate the 
LBL self assembled composite. Negatively charged $MnO_{2}$ and negatively charged 
(poly)sodium 4-styrenesulfonate (PSS) functionalized RGO was attached with positively charged PDDA
to fabricate 3d multilayer nanostructures
\newline
(PDDA/PSS-RGO/PDDA/$MnO_{2}$)\cite{lijmc}.
%%%%%%%%%%%%%%%%%%%%%%%%%%%%%%%%%%%%%%
\par
\textbf{(b) In-situ hybridization}\\
Achieving uniform surface coverage of nanocrystals on RGO is very difficult in the ex-situ method of
composite synthesis. In comparison, the in-situ technique gave uniform surface coverage by 
controlling the nucleation sites on RGO through surface modification.\\
Out of several possibilities, chemical reduction method has been the most popular technique to grow 
metal nanostructures on graphene. Noble metals' precursors, e.g. $AgNO_{3}$, $K_{2}PtCl_{4}$, 
$H_{2}PdCl_{6}$ can be easily reduced in-situ by chemical reducing agents e.g. sodium borohydride 
($NaBH_{4}$), amines and ascorbic acid. Au nanoparticle/graphene composite was synthesized by reducing 
$HAuCl_{4}$ with $NaBH_{4}$ in a octadecylamine solution of RGO\cite{muszjpcc}.
Along with nanoparticles, anisotropic nanostructures of metals on graphene have also been synthesized
using the method of chemical reduction. In the latter, self directing surfactants have been used to
get desired nanostructures on graphene.
For example, Au nanorods were prepared on seed modified RGO, by using cetyl trimethylammonium bromide 
(CTAB) as a shape directed surfactant\cite{kimlm} or 2.4 nm thick hexagonal close packed 
structured Au nanosheets on GO\cite{huangnc}. Photochemical reduction method has also been applied 
for the synthesis of nanostructured metal/graphene composite. In this technique, fluorescent gold 
nanodots on thiol modified RGO surface were grown by in-situ reduction of $HAuCl_{4}$ under light 
irradiation\cite{huangsml}.\\
In this in-situ reaction method, apart from the metal particles, metal oxides, such as, $SnO_{2}$
\cite{wangacsn}, $MnO_{2}$\cite{chenacsn} have also been synthesized on RGO surface. 
In-situ microwave irradiation technique was used to prepare metal nanoparticles as well as metal
oxides, on RGO, such as Cu,  Au, Ag\cite{hassanjmc} or $Co_{3}O_{4}$\cite{yanea}. 
In spite of large scale production, this method does not give good control over size and distribution
of the nanoparticles on RGO surface. This problem has recently been overcome by combining microwave
irradiation technique with ionic liquid assisted dispersion of RGO. In this modified method, 
Ru/RGO and Rh/RGO composites with narrow size distribution of Ru and Rh nanoparticles were 
synthesized\cite{marcarb}. \\
Recently, a novel, green approach, viz; electrochemical approach was used to fabricate 
graphene/inorganic composite. This method consisted of two steps; graphene sheets were first 
assembled on electrodes, which was followed by the immersion of graphene coated electrode in the 
electrolyte solution of the metallic precursors, for the electrochemical synthesis. 
Several high purity noble metals such as gold\cite{dujmc,fucpl} platinum\cite{liujps} and alloy of 
gold/platinum\cite{hupccp} were reduced electrochemically on graphene, by applying an 
electric potential. Apart from the metals, metal oxides were also synthesized with graphene using this 
method. P-type and n-type $Cu_{2}O$ films and ZnO nanorods were synthesized on 
(poly) ethylene tetraphthalate supported RGO films\cite{wujpcc}. In case of the $Cu_{2}O$ deposition 
the reduction of $Cu^{2+}$ and the pH of the electrolyte solution determined whether the semiconductor 
would be p-type or n-type. The chlorine doped n-type $Cu_{2}O$ on graphene was also synthesized by the 
same group of workers\cite{wujmc}.
\par
In-situ anodic electro-deposition of $\gamma-MnO_{2}$ nanoflowers on RGO electrodes was also 
carried out to fabricate $MnO_{2}$/RGO composite. For this RGO paper was first prepared by vacuum
filtration of the RGO solution followed by its use in desired shape as electrode for $MnO_{2}$ 
electro-deposition\cite{chengcarb}. \\
Apart from electrochemical depositions and electro-less deposition, which was previously observed 
for single walled carbon nanotube (SWCNTs) with Au and Pt nanoparticles\cite{choijacs}, GO/RGO has also
been used as templates for synthesizing Ag nanoparticles on them\cite{wujpcc}. Since the SWCNT has a 
higher Fermi level (less negative) than the redox potential of the metals Au or Pt, it act like cathode
to donate electrons for the reduction of the metal ions and its nucleation of the corresponding 
nanoparticles. For GO, the oxygen containing functional groups provide more nucleation sites for the 
nanoparticles rather than the RGOs, which are relatively free from the oxygen functionalities. 
Among the in-situ methods of preparing graphene/metal oxide composites, sol-gel method (which consists
essentially of hydrolysis and poly condensation reactions ) has also been used to prepare $TiO_{2}$
$SiO_{2}$, $Fe_{3}O_{4}$ nanostructures on RGO\cite{tangacsn,wangacsn,zhoucm}. The $OH^{-}$ groups
in the surface of GO/RGO play a key role and act as nucleation sites for the nanoparticles. Along 
with graphene/metal oxides composite fabrication by the sol-gel technique RGO/silica composite was 
also prepared by Ruoff and co-workers at Texas, by the hydrolysis of TMOS and water-ethanol dispersion 
of GO, which was then subsequently reduced to RGO by putting the composite film in a hydrazine hydrate 
vapor atmosphere overnight. The films showed a very low percolation threshold and behaved as a 
transparent conductor\cite{watchnl}. \\

Another green method for the synthesis of graphene/inorganic composite is hydrothermal synthesis
technique, in which high temperatures in a confined space generate high pressure and finally give
highly crystalline nanostructures, and also reduce GO to RGO. In this technique, CdS/RGO \cite{yecst}
was synthesized by using $Cd(Ac)_{2}$, GO, $Na_{2}S$ as precursors. The $Cd(Ac)_{2}$ was dissolved 
in the aqueous solution of GO and then  $Na_{2}S$ was added drop wise. The mixture was subjected to a
hydrothermal treatment in a Teflon coated autoclave shell at 453 K for 40h. Similarly, by using 
$Cd(CH_{3}COO)_{2}$ and GO in dimethyl sulfoxide (DMSO) in an autoclave at 453 K for 18h, CdS/RGO 
composite was synthesized\cite{caoadv}. The DMSO here acted as a source of sulfur and also a 
reducing agent to reduce GO to RGO.\\
Deposition technique is another one which is widely used to synthesize inorganic materials on different 
substrates. Physical vapor deposition, chemical vapor deposition and atomic layer deposition techniques 
were used to synthesize graphene/inorganic material composites.\\
PVD or physical vapor deposition technique is used to deposit material by condensation of the vaporized 
form of the material on a substrate. The technique can be of two types, viz; (a) vacuum deposition
and (b) sputtering deposition. The advantage of these methods is that they could use the graphene
(not the chemically modified one) with all its interesting properties, as the substrate. Gold (Au) 
film has been grown on micro-mechanically cleaved graphene layer at a rate of deposition
$1A^{o}s^{-1}$ under a pressure of $10^{-4}$Pa, and then subjecting it to a heat treatment at 1533 K.
The Au film morphology depended on the number of graphene layers and for lower number of layers, the 
particle size of Au nanoparticle was smaller and therefore, yielded high density\cite{zhoujacs}.
The theoretically predicted scaling law is given by $D\propto n^{1/3}$, where D is the mean particle
diameter and n is number of layers of the graphene film. Moir\'{e} pattern has also been observed for 
Ir, Pt, W, Re, on graphene/Ir(111)\cite{ndprl,njp11} and Pt, Pd, Rh, Co, Au on graphene/Ru(0001)
\cite{panapl,zhouss}.\\
%%%%%%%%%%%%%%%%%%%%%%%%%%%%%%%%%%%%%%%%%%%%%%%%%%%%%%%%%%%%%%%%%%%%%%%%%%%%%%%%%
CVD or chemical vapor deposition process has the same objective as the previous technique, but has some
advantages, e.g. inexpensive, controllable and high deposition rate, over PVD. This technique has 
also been employed to fabricate graphene supported CdSe\cite{yuacsami}, ZnO nanorod \cite{linsml}
and ZnS nanowires\cite{kimnt}.
The observed cathodoluminescence effect for ZnS/graphene composite has been described schematically in Fig. 4.
%%%%%%%%%%%%%%%%%%%%%%%%%%%%%%%%%%%%%%%%%%%%%%%%%%%%%%%%%%%%%%%%%%%%%%%%%%%%%%%%%%%%%%%%%
\begin{figure}
 \begin{center}
  \includegraphics[width=8.5cm]{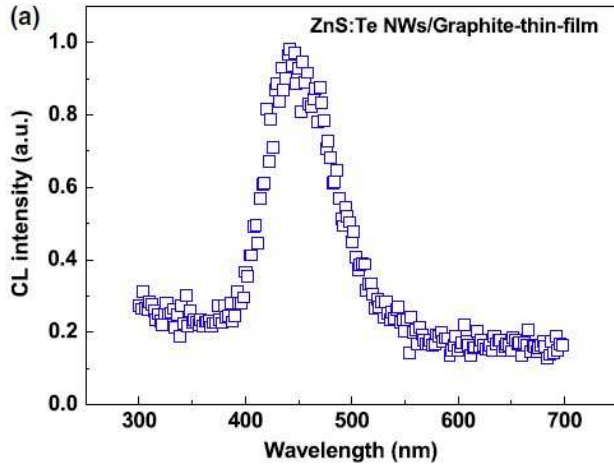}
 \end{center}
\caption{CL spectrum of ZnS:Te NWs grown on a MGF without
catalysts. Reproduced with permission from \cite{kimnt}, Institute of Physics Publishing.}
\end{figure}
%%%%%%%%%%%%%%%%%%%%%%%%%%%%%%%%%%%%%%%%%%%%%%%%%%%%%%%%%%%%%%%%%%%%%%%%%%%%%%%%%%%%%%%%

A representative example would be CdSe nanoparticle on RGO with
cadmium shot and selenium shot as precursors. GO was at first reduced to RGO by thermal annealing 
at 773K for 1h. under Ar flow. CdSe was prepared by CVD at 773 K under atmospheric pressure and RGO 
coated substrate was used to collect CdSe nanoparticles under a flow of Ar, which was used as a
carrier gas. Another useful material that has been fabricated using CVD is CNT/graphene composite 
in which CNT was grown vertically on graphene\cite{kondoape}. For this synthesis Ni\cite{zhangacsn} or Co
\cite{fanadv} was used as catalyst. The amount of these catalysts controlled the length and amount of
CNTs on graphene along with the deposition times. \\
Atomic layer deposition or ALD is different from the CVD process. In ALD, alternate and cyclic
supply of the gaseous precursors with sequential half reactions, allows the film to grow on the basis
of surface reactions. $SnO_{2}$ coated RGO has been synthesized by ALD by using $SnCl_{4}$ {\&}
$H_{2}O$ as precursors. By adjusting the growth temperatures to 473 K and 673 K respectively,
amorphous and crystalline $SnO_{2}$ nanoparticles were achieved by Meng and co-workers\cite{mengjpcc}. 
Also by adjusting the cycling numbers the morphology of the $SnO_{2}$ could be changed from nanoparticle 
to nanosheets.
%%%%%%%%%%%%%%%%%%%%%%%%%%%%%%%%%%%%%%%%%%%%%%%%%%
\section{Applications}
As discussed previously, graphene was successfully composited with different inorganic materials
as well as polymers and CNTs. The main objective for the preparation of composites is to use them 
for some practical applications. In this section we discuss about the various applicabilities of these
materials. \\
\subsection{Batteries and Ultracapacitors: energy storage}
Energy is the most important issue these days all around the world, and graphene composites may 
give a possible solution to this ever-increasing requirement. For low cost energy conversion and
storage material, lithium ion batteries (LiB) have no competitor these days. This is because of its
high potential with respect to standard hydrogen cell (-3.04 eV) and large energy density of theoretical
value 4000$Whkg^{-1}$\cite{palacincsr}. Graphene and its derivatives do not host Li like bulk graphite
, but they can store Li via surface adsorption, which then have high conductivity and surface area. 
So, different metal oxide nanostructures e.g. $SnO_{2}$\cite{wangacsn}, $Co_{3}O_{4}$
\cite{wuacsn}, $MnO_{2}$\cite{wangjacs3}, $Fe_{3}O_{4}$\cite{zhoucm} have been composited with graphene
for lithium ion battery applications. For bare metal oxides, the problem arises due to the rapid 
decay of the capacity, which is caused by poor conductivity of the materials. After being attached with 
graphene, the hybrid electrodes have excellent conducting network, as already observed\cite{wangjacs3}.
To explain, we can take the example of the $Mn_{3}O_{4}$/graphene anode. Being a low cost and high
capacity (theoretical value of $\sim 936 m A h g^{-1}$) system, $Mn_{3}O_{4}$ would be considered as 
a possible candidate for the anode material in LiB. But the low conductivity has limited its 
practical applications. The growth of $Mn_{3}O_{4}$ nanoparticles on graphene not only solves 
this problem of conductivity directly, but also provides a specific capacity value, which is close to
the theoretical one, and more than double that observed in pure $Mn_{3}O_{4}$. Also the graphene/metal
oxide composite provides a large surface area for the nanoparticles to grow, hence the agglomeration
of the nanoparticles is reduced. The specific capacity is enhanced in the composites due to this large
active surface of the nanoparticles. This large surface allows the nanoparticles to participate 
more efficiently in Li/electron diffusion, during discharge/charge cycles. It was also seen that the
structure and crystallinity of the nanoparticles remained unaltered after many charging and discharging cycles. 
Similarly, graphene coated metal oxide nanoparticles have also proved to be very good anode material
for LiBs. For example, graphene wrapped $Co_{3}O_{4}$ with very low 'C' content, provides excellent
capacity of $1000 m A h g^{-1}$ for more than 130 cycles\cite{yangac}. Bare graphene due to large 
specific surface area shows quite a high specific capacity $\sim 540 m A h g^{-1}$), the graphene 
composite with other carbon material, e.g. CNTs, increase the separation between graphene stacks and
hence increase the specific capacity by 40{\%}. \\
Apart from lithium ion batteries, ultracapacitor is another type of electrochemical energy storage 
device, which gives high power density, long cycling life, in comparison to conventional
battery device. By their operational mechanisms, ultra or supercapacitors are of two types; \\
(a) Electrical double layer capacitor (EDLC), which stores energy via electrostatic process, viz;
charge accumulation occurs at electrolyte/electrode interface due to polarization. EDLC thus requires
an electrode material with high conductivity and large surface area. The RGO is a promising candidate
since it provides an \textquotedblleft open pore \textquotedblright structure which allows the 
electrolyte ions to move in and form electric double layers\cite{liuadv}. \\
(b) Pseudo capacitor is based on the rapid redox reaction of the chemical species in the electrode. 
Usually, metal oxides and conducting polymers have been used in these electrodes. But the high cost
and relatively low conductivity of the metal oxides and conducting polymers, have limited their uses. 
Graphene derivatives may offer a solution to these problems. \\
Though graphene has been used as the electrode materials of supercapacitors\cite{cnrjcs}, 
graphene composited with different metal oxides
e.g. ZnO\cite{zhangjec}, $Co_{3}O_{4}$\cite{yanea}, $RuO_{2}$\cite{kimadvfm} have also been used to enhance the effect.
The metal oxide nanoparticles contribute to the energy storage and the RGO sheets provide the capacitance
by the electron double layer mechanism at the carbon surface. The RGOs also create the conducting
network for the nanoparticles. For example, $MnO_{2}$/graphene composite electrode showed high specific 
capacitance ($\sim 310 F g^{-1} at 2 mV s^{-1}$), which was 3 times higher than that observed for 
pure RGO or $MnO_{2}$. Cheng and others \cite{wuacsn} had worked on graphene/metal oxide hybrid
electrode, to improve the energy density of the supercapacitor while maintaining its high power density.
They had used graphene as negative electrode and $MnO_{2}$ nanowire/graphene composite as positive 
electrode and observed high energy density, which is the square of the operating voltage 
across the cell. This asymmetric capacitor system showed much higher 
energy density ($\sim 30.4 W h kg^{-1}$) in comparison to the symmetric one. \\
Apart from metal oxide /graphene composites, graphene/conducting polymer composites were also used as 
supercapacitors.
They have moderate conductivity and fast charge-discharge kinetics. Also the flexibility for thin 
film based electronics, added a new interest to them. (Poly) aniline has been used mostly with
RGO/GO to be used as supercapacitors\cite{ramjpsb}. It has been reported that 1{\%} GO loading 
in the backbone of (poly) aniline gives an enhanced conductivity of $10 S cm^{-1}$ and specific 
capacitance of $531 F g^{-1}$, whereas for pure (poly) aniline the conductivity and
specific capacitance are reported to be $2 S cm^{-1}$ and $260 F g^{-1}$ respectively\cite{ramjpsb}.
The cycling stability was also 
enhanced in comparison to pure (poly) aniline as the GO underwent the mechanical deformation during
charging-discharging cycles\cite{wangacsami2}. \\
\subsection{Fuel cell and solar cell: energy generation}
Along with the storage of energy (as done by batteries and ultracapacitor) the generation of energy is
also a problem of interest. Fuel cells generate electricity by the reaction between 
fuel and oxidant, which act as anode and cathode respectively. Starting with hydrogen/oxygen fuel cells,
a number of different combinations have been tried so far for the fuel cell fabrication. One of the
most popular and widely used materials for low temperature fuel cells is platinum. But the problem 
with Pt-based catalysts is that Pt is very costly, hence its uses should be minimized, and on the 
other hand, decreasing Pt content hampers the performance of the fuel cell. \\
Large effective area of graphene makes it a possible candidate for fuel cell applications. Pt/Graphene
composite has been used in methanol oxidation cells\cite{dongcarb,licarb} and oxygen reduction cells
\cite{jafrijmc,shaojps}. The 2D structure of graphene makes both sides of the material exposed to the 
solution, hence increases the effective surface area, which enhances the efficiency. The residual
oxygen functionality (however small) of RGO removes the carbonaceous species and hence it shows 
better tolerance to carbon monoxides, which is evolved during methanol oxidation\cite{licarb}. Also the 
nitrogen doped graphene, with higher conductivity has been used for the fuel cell 
application\cite{zhangpccp}. This enhanced electronic conductivity essentially increases the 
electrocatalytic activities of N-RGO-Pt catalysts. In methanol fuel cell, these exhibit oxidation 
current of $135 mA mg^{-1}$, which is almost double that of  bare RGO-Pt catalyst.
\par
For solar cell applications, amorphous carbon has been used for several years, as p-type semiconductor
with silicon as n-type semiconductor\cite{masemsc}. Apart from this, CNT has  also been employed 
with silicon heterojunction for the solar cell application\cite{jiaadv}. But both of these had several 
drawbacks, which did not allow for wide usage. For amorphous carbon the tuning of the
electronic properties was the main obstacle, whereas for the CNTs the bundling of the nanotubes,
reduced the film connectivity and conductivity significantly. Graphene, on the other hand has a large 
surface continuity and its properties can be tuned by doping and functionalization. The graphene sheets
of $100 \mu m$ size, on n-Si with 100{\%} coverage (prepared in CVD method) have been used 
to make Schottky junction solar cell\cite{liadv}, which showed 1.5{\%} efficiency with a fill factor
(obtained maximum power/ theoretical power) of 56{\%}. The graphene film behaves as a semi-transparent
electrode and generate a built-in voltage of 0.55 V for the electron-hole separation\cite{liadv}.
Although the graphene based solar cell systems show reasonable value of efficiency, it is still less than that of
pure Si-based solar cells. Hence, increasing of the efficiency is one of the major aspects of graphene
based solar cell devices, where graphene-based composites may give better results. The graphene based polymer composites
have been incorporated as electrodes, electron and hole transports for photovoltaic device
applications \cite{yinsml,yuacsn,liacsn}.
%%%%%%%%%%%%%%%%%%%%%%%%%%%%%%%%%%%%%%%%%%%%%%%%%%%%
\section{Conclusions and future directions}
The unique electronic, mechanical and thermal properties of graphene and its derivatives
along with its low cost mass production make it a promising material for composites with
different polymers, metals, metal oxides, and other carbon based materials. \\
Starting with simple solution mixing to CVD synthesis
a number of methods have been employed to fabricate 
the graphene composite for several application purposes. Some of the investigations
on graphene and graphene based composites  have been highlighted in
this review article. We have discussed the most important of today's scientific challenge,
\textquoteleft The energy generation and its storage \textquoteright
and its possible solutions from the point of view of
graphene composites. With a large number of new research being reported
and it would not be an overestimation,to expect several new fabrication techniques  to evolve.
Harvesting of solar energy by the graphene
composite has not been achieved in laboratories. So it would be one of the key problems for the 
next decade or so. Also the storage of hydrogen has been observed in graphene\cite{cnrpnas2}, it would be profitable
to explore  the possibility of increase the storage capacity by suitable graphene composites. 
%%%%%%%%%%%%%%%%%%%%%%%%%%%%%%%%%%%%%%%%%%%%%%%%%%%%%%%%%%%%%%%%%%%%%%%%%%%%%
\section*{Acknowledgements}
The authors thank Department of Science and Technology, New Delhi, India for the funding through an
Indo-Australian project on 'Nanocomposite and Clean Energy'. Sreemanta Mitra thanks University grants 
Commission, New Delhi, India, for Senior research fellowship and Dipankar Chakravorty thanks 
Indian National Science Academy for Honorary Scientist's position.
%%%%%%%%%%%%%%%%%%%%%%%%%%%%%%%%%%%%%%%%%%%%%%%%%%%%%%%%%%%%%%%%%
%Merlin.mbs v4.21 2009-07-09.
%

%\bibliography{bibliographyrev}
%}
\end{document}